\documentclass[amsmath,amssymb,aps,prl,reprint,floatfix]{revtex4-1}

\usepackage{graphicx}% Include figure files
\usepackage{dcolumn}% Align table columns on decimal point
\usepackage{bm}% bold math
\usepackage[none]{hyphenat}
\usepackage[mathlines]{lineno}% Enable numbering of text and display math
\usepackage{float}
\usepackage [english]{babel}
\usepackage [autostyle, english = american]{csquotes}
\MakeOuterQuote{"}

\begin{document}

\title{Robust entanglement gates for trapped-ion qubits}% Force line breaks with \\

\author{Yotam Shapira}
\email{yotam.shapira@weizmann.ac.il}
\author{Ravid Shaniv}
\author{Tom Manovitz}
\author{Nitzan Akerman}
\author{Roee Ozeri}
\affiliation{%
Department of Physics of Complex systems, Weizmann Institute of Science, Rehovot 7610001, Israel
}%

\date{\today}% It is always \today, today,

\begin{abstract}
High-fidelity two-qubit entangling gates play an important role in many quantum information processing tasks and are a necessary building block for constructing a universal quantum computer. Such high-fidelity gates have been demonstrated on trapped-ion qubits, however, control errors and noise in gate parameters may still lead to reduced fidelity. Here we propose and demonstrate a general family of two-qubit entangling gates which are robust to different sources of noise and control errors. These gates generalize the celebrated M\o lmer-S\o rensen gate by using multi-tone drives. We experimentally implemented several of the proposed gates on $^{88}\text{Sr}^{+}$ ions trapped in a linear Paul trap, and verified their resilience.
\end{abstract}

\pacs{Valid PACS appear here}% PACS, the Physics and Astronomy
\maketitle
Quantum information processing (QIP) is a rapidly growing field. Two-qubit entanglement gates play an essential part in many QIP tasks, and in particular serve as part of a universal gate set \cite{Barenco1995,Kitaev1997}. Moreover, the ability to execute high-fidelity gates is necessary in order to achieve fault-tolerant quantum computing \cite{Aharonov2008}. To this end, gates which are less sensitive to control errors would be beneficial.

Trapped-ions are a promising platform for QIP applications, and therefore implementing high-fidelity two-qubit entangling gates in trapped-ion systems has been at the center of many experimental investigations  \cite{Leibfried2003,Kirchmair2009,Gaebler2016,Ballance2016}. In most of these demonstrations however fine-tuning and stability of gate parameters were required. Recently several gates were demonstrated with increased resilience against different noises \cite{Roos2008,Haddadfarshi2016,Palmero2017,Schafer2017,Leung2017,Manovitz2017}.

Here we present a general scheme for engineering robust entanglement gates in trapped ion systems. We generalize the well-known M\o lmer-S\o rensen (MS) gate \cite{Molmer1999,Sorensen2000} by using additional frequency components in the laser drive. We show that these components may be used as additional degrees of freedom to optimize gate robustness to different noise processes  and control errors. Our main results are entangling gates that are robust to gate-timing errors, harmonic trap frequency uncertainties and off-resonance couplings to neighbouring transitions.

We experimentally demonstrate our gates on a $^{88}\text{Sr}^+$, two-ion crystal, trapped in a linear Paul trap. We define the states $\left|S\right\rangle\equiv5S_{\frac{1}{2},\text{-}\frac{1}{2}}$ and $\left|D\right\rangle\equiv4D_{\frac{5}{2},\text{-}\frac{3}{2}}$ as our qubit levels. The $\left|S\right\rangle\leftrightarrow\left|D\right\rangle$ optical quadrupole transition at $674\text{ nm}$ is driven by a narrow linewidth ($<20\text{ Hz}$) laser. We cool the ion-crystal normal mode to an average phonon number $\bar{n}\approx0.17$, although some of our data was taken in a Doppler cooled regime with $\bar{n}\approx10$ for comparison. More details about our system can be found in  \cite{Akerman2012,Akerman2015}. 

The MS entangling gate is implemented by driving the qubit transition in a two-ion crystal with a bi-chromatic field containing the frequencies $\omega_{\pm}=\omega_{0}\pm\left(\nu+\xi_0\right)$. $\omega_{0}$ is the resonance frequency of the qubit transition, $\nu$ is the harmonic trap frequency of a selected normal mode of the crystal and $\xi_0$ is a frequency detuning from the side-band transition. Following \cite{Sorensen2000}, the effective interaction Hamiltonian is: $\hat{H}=-2\hbar\eta\Omega\hat{J}_{y}\left(\hat{a}e^{i\xi_{0}t}+h.c\right)$, where $\hat{a}$ is the mode lowering operator, $\eta$ is the Lamb-Dicke parameter, $\Omega$ is the laser's Rabi frequency and $\hat{J}_{y}=\frac{\boldsymbol{1}\otimes\hat{\sigma}_{y}+\hat{\sigma}_{y}\otimes\boldsymbol{1}}{2}$ is a global $\hat{y}$ rotation operator in the two-qubit subspace. This interaction yields the unitary evolution operator,
\begin{equation}
\hat{U}\left(t;0\right)=e^{-iA\left(t\right)\hat{J}_{y}^{2}}e^{-iF\left(t\right)\hat{J}_{y}\hat{x}}e^{-iG\left(t\right)\hat{J}_{y}\hat{p}},\label{MS-U}
\end{equation}
with:
\begin{equation}
\begin{cases}
F\left(t\right)=-\frac{\sqrt{2}\eta\Omega}{\xi_{0}}\left(\sin\left(\xi_{0}t\right)\right)\\
G\left(t\right)=\frac{\sqrt{2}\eta\Omega}{\xi_{0}}\left(1-\cos\left(\xi_{0}t)\right)\right)\\
A\left(t\right)=-\int\limits _{0}^{t}d\tau F\left(\tau\right)\partial_{\tau}G\left(\tau\right)
\end{cases}.\label{MS-FGA}    
\end{equation}
Eq. \ref{MS-U} implies that the ions' evolution follows a trajectory $\left(G\left(t\right),F\left(t\right)\right)$ in the normal mode's phase space, which depends on the eigenvalue of $\hat{J}_{y}$. In the $\hat{J}_{y}$ basis it acquires a geometric phase $A\left(t\right)$, which corresponds to a correlated rotation by an angle equal to the area enclosed by the trajectory. By choosing $\xi_0=2\eta\Omega$ the phase space trajectory closes at the gate time, defined by $T\equiv\frac{\pi}{\eta\Omega}$ such that $A(T)=\frac{\pi}{2}$. 

Ideally the MS gate generates deterministic, temperature-independent, two-qubit entanglement. However this requires exact calibration and short-term stability of the gate parameters, such as the gate time $T$ and the normal mode frequency $\nu$. Inaccuracies and drifts will result in a reduced, temperature-dependent, gate fidelity. Furthermore the MS scheme is substantially slower than the normal mode period in order to avoid unwanted direct carrier coupling.

To mitigate these shortcomings we generalize this entangling scheme by employing additional frequency tones. The amplitudes of these tones are then treated as additional degrees of freedom and used to reduce, and in some cases eliminate, the effect of errors in different gate parameters on its fidelity, leading to robust entanglement. Here we demonstrate robustness to gate timing-errors, normal mode frequency errors and the reduction of direct carrier coupling.

Our generalized entangling gate is formally implemented by driving the ions with a multi-chromatic beam containing the frequencies $\omega_{\pm,i}=\omega_{0}\pm\left(\nu+n_i\xi_0\right)$ with relative amplitudes $r_i$ and phases satisfying $\phi_{+,i}+\phi_{\text{-},i}=0$. We interpret $r_i>0$ ($<0$)  as $\phi_{+,i}-\phi_{\text{-},i}=0$ ($=\pi$). Here the $n_i$'s are a set of $N$ parameters, where $N$ is the number of tones.

We now turn to derive the generalized gate Hamiltonian. We assume the ions are well within the Lamb-Dicke regime and we may expand the interaction Hamiltonian in orders of $\eta$ to obtain,
\begin{equation}
    \hat{H}=2\hbar\Omega\hat{J}_{x}e^{i\nu t}-2\hbar\eta\Omega\hat{J}_{y}\sum_{i=1}^{N}\left(\hat{a}e^{i\xi_{0}n_{i}t}+h.c\right).\label{IntHam}
\end{equation} 
The first term in Eq. \ref{IntHam} generates unwanted off-resonance carrier coupling, i.e local qubit excitations without involving the normal mode. By assuming that $\Omega\ll\nu$ this term may be neglected in a rotating wave approximation.
Since the commutation relation of any of the remaining terms in Eq. \ref{IntHam} are proportional to the identity, the Hamiltonian can be exactly solved. The resulting unitary operator is given by the same expression as in Eq. \ref{MS-U}, however with a generalized trajectory given by,
\begin{equation}
\begin{cases}
F\left(t\right)=-\frac{\sqrt{2}\eta\Omega}{\xi_{0}}\sum\limits_{i=1}^{N}\frac{r_{i}}{n_{i}}\left(\sin\left(n_i\xi_{0}t\right)\right)\\
G\left(t\right)=\frac{\sqrt{2}\eta\Omega}{\xi_{0}}\sum\limits_{i=1}^{N}\frac{r_{i}}{n_{i}}\left(1-\cos\left(n_i\xi_{0}t)\right)\right)
\end{cases}.\label{FGA}    
\end{equation}

Once again, the normal-mode motion follows a trajectory $\left(G\left(t\right),F\left(t\right)\right)$ in phase space, accompanied by a correlated rotation in the two-qubit subspace by an angle corresponding to the area enclosed by the trajectory.

Similarly to the MS case we demand that $G\left(T\right)=F\left(T\right)=0$ and $A\left(T\right)=\frac{\pi}{2}$, where $T$ is the gate time. This condition is satisfied by setting, 
\begin{equation}
n_{i}\in\mathbb{Z}\text{   ;   }\sum\limits_{i=1}^{N}\frac{r_{i}^{2}}{n_{i}}=1\text{   ;   }\xi_{0}=2\eta\Omega=\frac{2\pi}{T}.\label{Cond}
\end{equation}
Eq. \ref{Cond} defines a family of entangling gates that differ by the harmonic tones used $\left\{ n_{i}\right\}_{i=1}^{N}$ and their relative amplitudes $\left\{r_{i}\right\}_{i=1}^{N}$. The case $N=1$, $n_1=r_1=1$ retrieves the usual MS entangling scheme. These parameters can be used as additional degrees of freedom for optimizing the entangling gate's robustness to different noises or control errors.

To see how gate robustness can be increased we calculate the gate fidelity, $F_{g}\equiv\left\langle \psi_{\text{ideal}}\left|\hat{\rho}\left(T\right)\right|\psi_{\text{ideal}}\right\rangle$, where $\psi_{\text{ideal}}$ is the ideal two-qubit final state and $\hat{\rho}$ is the two-qubit density matrix during the gate evolution. Here, $\hat{\rho}$ is obtained by tracing out the normal mode degrees of freedom, yielding,
\begin{equation}
F_{g}=\frac{3+e^{-4\left(\bar{n}+\frac{1}{2}\right)\frac{F^{2}+G^{2}}{2}}}{8}+\frac{e^{-\left(\bar{n}+\frac{1}{2}\right)\frac{F^{2}+G^{2}}{2}}\sin\left(A+\frac{FG}{2}\right)}{2}.\label{Fidelity}
\end{equation}
Where $\bar{n}$ is the average phonon number assuming an initial thermal state. $F_{g}$ depends on various physical parameters such as $\xi_0$, $\nu$ and $T$. The closed form of Eq. \ref{Fidelity} allows for the expansion of $F_{g}$ in any parameter of choice, $\alpha$, which may deviate from its ideal value $\alpha_0$, in the experiment due to imperfections,
\begin{equation}
F_{g}\left(\delta\alpha\right)=1+\frac{\partial^{2}\left(F_{g}^{2}\right)}{\alpha^{2}}|_{\alpha_{0}}\left(\delta\alpha\right)^{2}+\mathcal{O}\left(\delta\alpha^4\right).\label{FidExp}  
\end{equation}
Increasing robustness to deviations in $\alpha$ is possible if the leading order contribution to the fidelity may be minimized or even eliminated, by a specific choice of $\left\{n_{i}\right\}$ and $\left\{r_{i}\right\}$. 

Here we show that elimination of errors order-by-order is indeed possible for gate timing-errors $T=T_{0}+\delta T$ and reducing the leading order contribution is possible for normal mode frequency errors $\nu=\nu_{0}+\delta\nu$. In addition we show that increasing the robustness to gate timing-errors immediately reduces infidelities due to direct off-resonance carrier coupling, which determines the shortest possible gate time under a certain error.

\begin{figure}
\includegraphics[width=8.5cm]{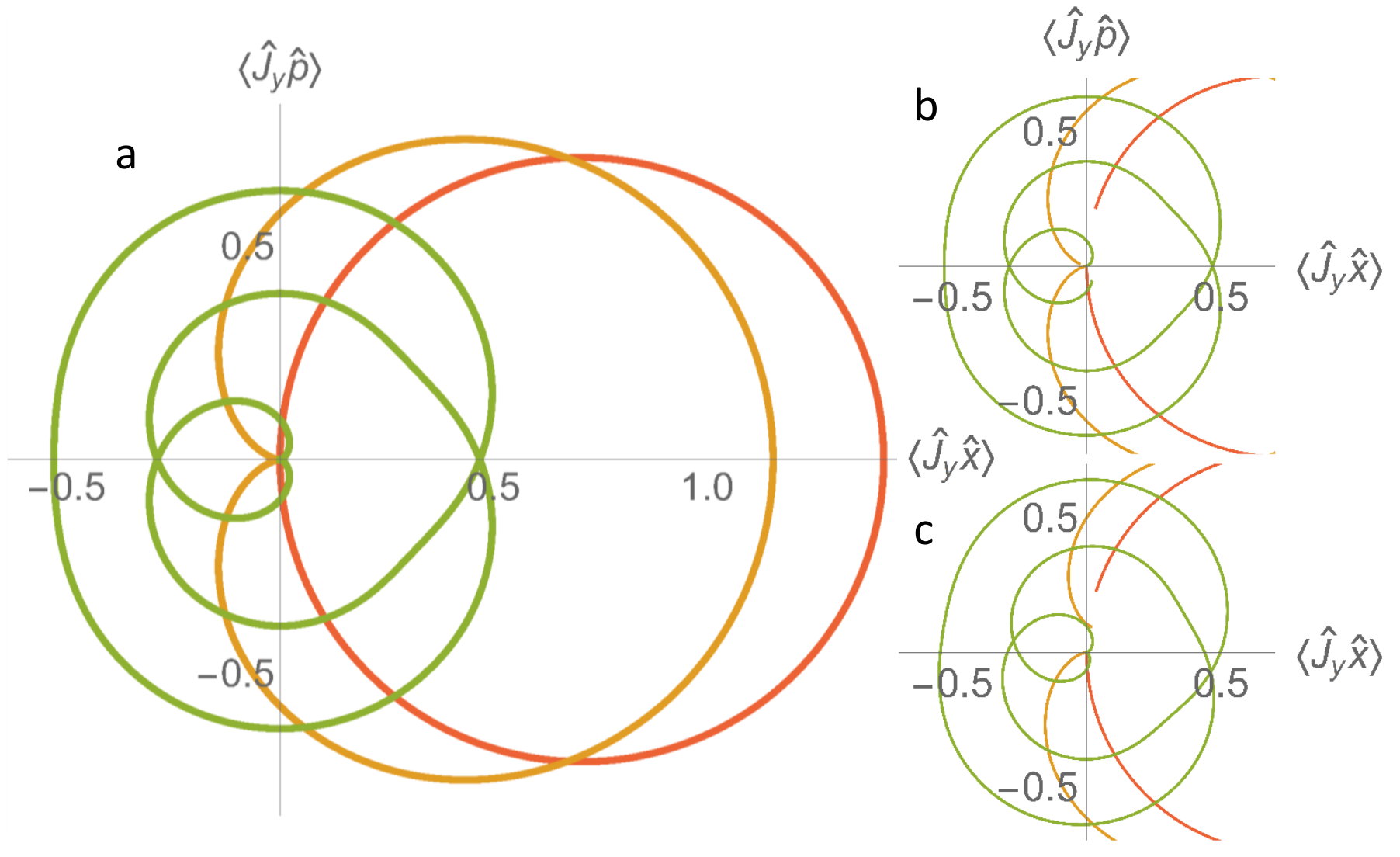}
\caption{\textbf{Entangling gate trajectories in phase space.} \textbf{(a)} Ideal case showing the MS gate (red), the heart-shaped Cardioid(1,2) (orange) and the CarNu(2,3,7) (green). Without errors all gates start and end at $\left(0,0\right)$ and enclose an area of $\frac{\pi}{2}$. \textbf{(b)} Same gates with a $5\%$ gate time error, the MS gate (red) no longer closes the trajectory, while the other gates remain significantly closer to the origin. \textbf{(c)} Same gates with a $5\%$ normal mode frequency error, the CarNu(2,3,7) gate is almost unaffected while the other gates fail to close the trajectory.}\label{FigPhase}
\end{figure}

The first optimization procedure we demonstrate is that of gate timing-errors. We set $T=T_0+\delta T$ and expand the fidelity function. Eliminating the first $N-1$ leading orders in $\delta T$ reduces to a set of constraints $V_{\boldsymbol{n}}\boldsymbol{r}=\boldsymbol{0}$ where $\boldsymbol{r}$ is a vector of $N$ amplitudes and $V_{\boldsymbol{n}}$ is a $N\times N$ Vandermonde matrix defined by $\left(V_{\boldsymbol{n}}\right)_{i,j}=\left(n_{j}\right)^{i-1}$. There are no constraints on the $n_i$'s. Satisfying these leads to a fidelity of the form $1-F_{g}\sim\left(\frac{\delta T}{T}\right)^{2N}$. For $N=2$, i.e. eliminating the 2nd order term, our solution yields $r_1=-r_2$, and the gate time is minimized by choosing $n_1=1$ and $n_2=2$. With this choice we obtain an entangling gate with a Cardioid (heart-shaped) phase space trajectory. Inspired by this, we denote all entangling gates which satisfy this set of constraints as \textit{Cardioid gates}. Specifically, the heart-shaped gate is then a Cardioid(1,2). Fig. \ref{FigPhase}a  shows a comparison between the phase space trajectory formed by the MS (red line) and Cardioid(1,2) (yellow line) gates. Fig. \ref{FigPhase}b shows the same corresponding trajectories in the presence of a gate-timing error.

The reduction of gate error in the Cardiod gate is the result of coherent interference of the contributions from the two amplitudes, $r_1$ and $r_2$. To show this we compare the Cardioid with an entangling gate which uses the same frequency tones and amplitudes only this time all amplitudes have the same sign. The resulting gate has an increased sensitivity to gate-timing errors. Since this gate has the opposite effect on the sensitivity to timing errors we refer to this family of gates as \textit{Antioid gates}. 

\begin{figure}\includegraphics[width=8.5cm]{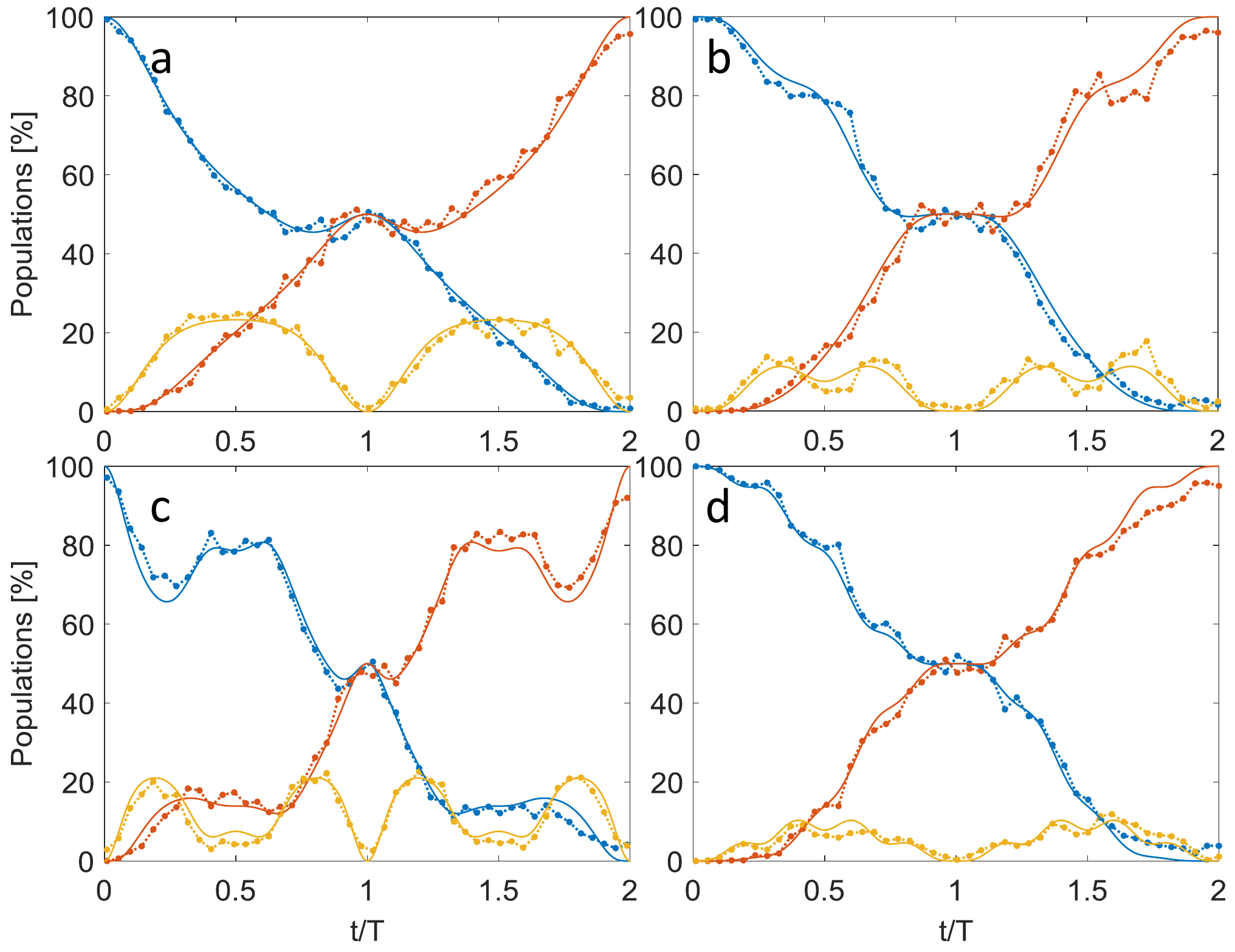}
\caption{\textbf{Population evolution of the different entangling gates.} All data points in this and following figures were obtained by averaging over 625 realizations yielding $2\%$ projection error at $50\%$ probability. Data points are shown by connected filled circles and an analytic solution, with no fitting parameters, is shown by solid lines. \textbf{(a)} MS gate. shown as a benchmark for the other driving schemes. \textbf{(b)} Cardioid(2,3) gate. The plot shows a flat response of the two-qubit state populations around the gate time. \textbf{(c)} Antioid(2,3) gate. Here we use the same tones and power as in the Cardioid(2,3) gate only with opposite phase. We see a narrow quadratic response around the gate time. \textbf{(d)} CarNu(2,3,7) gate. The scan has the same gate time error response scaling as Cardioid(2,3) gate, in addition it is also less sensitive to normal mode frequency errors.\label{FigTime}}\end{figure}

Fig. \ref{FigTime}a, \ref{FigTime}b and \ref{FigTime}c show the measured and analytically calculated time evolution of the qubit state population for the MS, Cardioid(2,3) and Antioid(2,3) gates respectively \footnote{Non-linear response of system components generates unwanted additional tones. As an example for $n_1=1$ and $n_2=2$, a third order non-linearity generates on-resonance side-band drive. Thus we mostly used tones which have no on-resonance third order response, e.g. $n_i\in\left\{ 2,3,7,8,12,...\right\}$.} . Compared to the standard MS gate, the Cardiod(2,3) shows a variation of the ion state populations around the gate time which is flatter than quadratic, while the Antioid(2,3) has a faster quadratic response. Furthermore, by looking at the SD+DS populations, (orange filled-circles - data; orange solid line - analytic solution) which are correlated with excitation of motion, it can be seen that in the Cardiod case, less motion is excited and the ion is closer to the phase-space origin throughout the gate (see also Fig. \ref{FigPhase}). 

\begin{figure}\includegraphics[width=8.5cm]{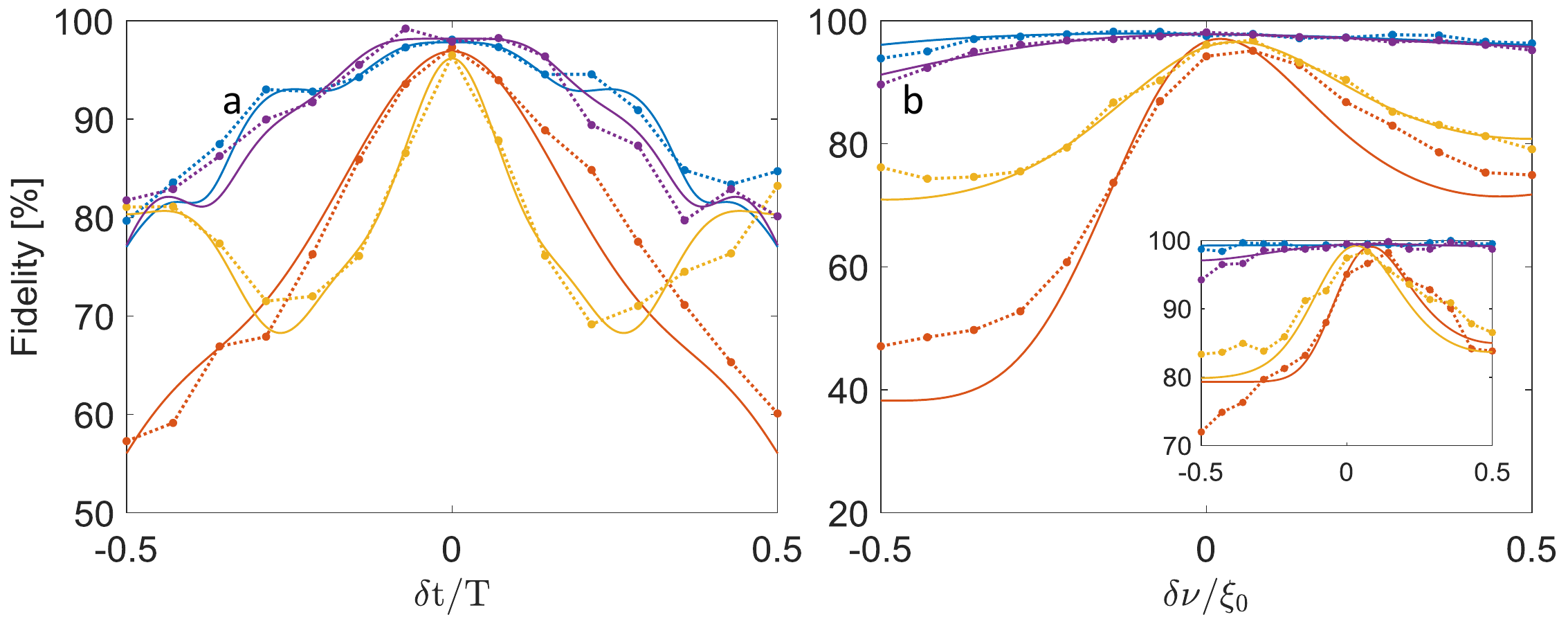}
\caption{\textbf{Gate fidelities.} \textbf{(a)} Gate fidelity vs. timing error. The MS gate (red) and the Antioid(2,3) gate (orange) have a quadratic fidelity dependence on the gate time error. The CarNu(2,3,7) gate (blue) and the Cardioid(2,3,7) gate have a 4th order and a 6th order fidelity dependence on the gate time error respectively, and are therefore more resilient. \textbf{(b)}  Gate fidelity vs. harmonic trap frequency errors. Here optimization is not order-by-order. Rather, we minimize the leading order quadratic pre-factor. The CarNu(2,3,7) gate (blue) shows the flattest response compared to the other driving schemes. The inset shows the gate purity, where the optimization allows eliminating the leading order for the CarNu(2,3,7).\label{FigFidel}}\end{figure}

Fig \ref{FigFidel}a compares the gate fidelity of the Cardioid(2,3), Antioid(2,3) and MS gates for different gate times. As seen, the Cardioid gate demonstrates a response to timing errors which scales as $(\frac{\delta T}{T})^4$ while the Antioid and MS gates have a quadratic response, which is narrower for the Antioid.  

\begin{figure}\includegraphics[width=8.5cm]{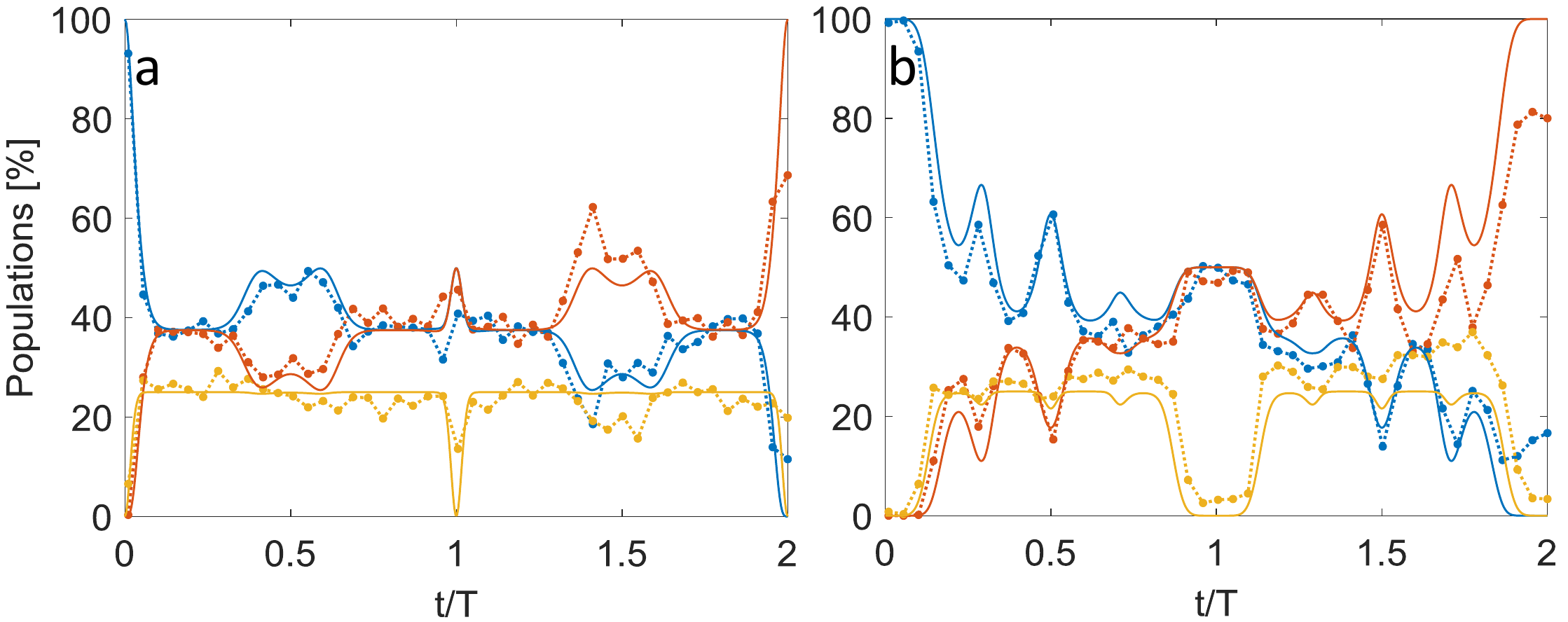}
\caption{\textbf{Finite temperature gates.} Here ions are Doppler cooled to $\bar{n}\approx9.8$ \textbf{(a)} Antioid(2,3) gate population evolution. A narrow change around $t=T$ marks ion entanglement. Gate fidelity is very sensitive to timing-errors. \textbf{(b)} Cardioid(2,3,7,8) gate population evolution. Using four tones increases the robustness resulting in a wide feature around gate time.}\label{FigHot}\end{figure}

Due to the dependence of the fidelity on $\bar{n}$ in Eq. \ref{Fidelity}, any small deviation of $F\left(T\right)$ or $G\left(T\right)$ from $0$ will result in errors that are exponentially amplified by $\bar{n}$. Therefore robustness is especially important when implementing entangling gates in systems which are not cooled to the normal mode ground state. Fig. \ref{FigHot} shows population evolution of different gates and their corresponding robustness to timing-errors in a Doppler cooled regime, with $\bar{n}\approx9.8$. As seen the Antioid gate is highly sensitive to gate-time errors at this high temperature, whereas the Cardioid gate shows a flatter time response.

An intuitive understanding of the origin of robustness to time errors can be gained by observing that the Cardioid constraint is simply a demand for order $N$ smoothness of $F$ and $G$ at the beginning and end of the gate. This means that the spin dependent forces applied by the gate vary smoothly towards the gate beginning and end as well. The use of a smooth gate envelope (e.g. Gaussian) was shown before to be beneficial for reducing off-resonance carrier coupling, generated by the fast rotating term in Eq. \ref{IntHam}. 

\begin{figure}\includegraphics[width=8.5cm]{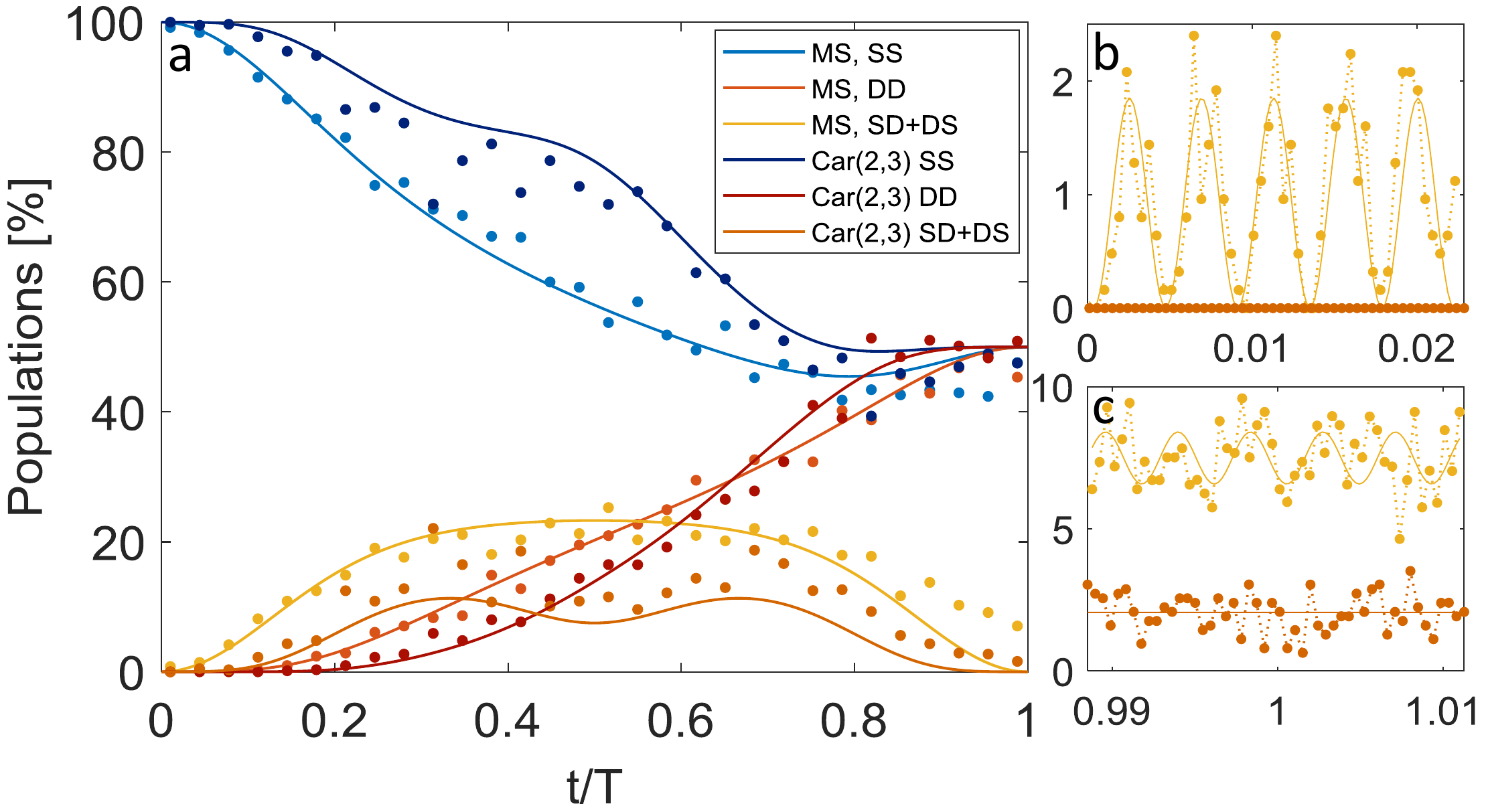}
\caption{\textbf{Off-resonance carrier coupling.} \textbf{(a)} Population evolution of the MS and Cardioid(2,3) gates. Here $\frac{\Omega}{\nu}\sim5\%$ which corresponds to $2\%$ infidelity for the MS gate and $0.1\%$ for the Cardiod(2,3) gate. This infidelity is seen by the SD and DS populations (yellow for MS and orange for Cardioid) not nulling by the end of the gate. \textbf{(b)} Zoom-in scan of SD+DS populations at the gate start. Fast carrier oscillations are easily observed for the MS gate (yellow points). These are heuristically fitted to off-resonance Rabi oscillations $A \sin^{2}\left(\tilde{\Omega}t\right)$ with an effective Rabi frequency $\tilde{\Omega}=\sqrt{\Omega^{2}+\nu^{2}}$ (yellow line). For the Cardioid no such oscillations are seen. \textbf{(c)} Zoom-in scan of SD+DS populations around gate time. The carrier oscillations are less distinct since here projection noise is on par with the oscillation amplitude. The MS gate populations (yellow points) oscillate around $8\%$ at gate time, while the Cardioid populations oscillate around $2\%$, which corresponds directly to an increased gate fidelity. The $2\%$ Cardioid infidelity is due to other imperfections and not carrier coupling.}\label{FigOff}\end{figure}

We next turn to optimizing our gate to mitigate off-resonance carrier coupling. The gate error due to off-resonance carrier coupling is given by \cite{Sorensen2000},
\begin{equation}
1-F_{d}=\sum_{n=0}^{\infty}\frac{\left(-1\right)^{n}}{\left(2n\right)!}\left(\frac{2\Omega\sin\left(\nu T\right)}{\nu}\sum_{i=1}^{N}r_{i}\sum_{j=0}^{\infty}\left(\frac{2n_{i}\eta\Omega}{\nu}\right)^{j}\right)^{2n}.\label{EqnInfd}
\end{equation}
Interestingly, eliminating this term in increasing orders of $\frac{\Omega}{\nu}$ yields exactly the Cardioid set of constraints, indicating that the Cardiod gate will eliminate both timing as well as carrier coupling errors. As an example, a typical MS gate with $\frac{\Omega_0}{\nu}=0.1$ and gate time $T_0$ suffers from $\sim2\%$ infidelity due to off-resonance carrier coupling. Increasing the Rabi frequency to $3\Omega_0$ will result in a gate time $\frac{1}{3}T_0$ but also increases the infidelity by an order of magnitude. However using a Cardioid(1,2) gate with a $3\Omega_0$ Rabi frequency will result in $\sim0.1\%$ infidelity with a gate time $\sqrt{\frac{8}{9}}T_0\approx0.55T_0$. Fig. \ref{FigOff} shows the effect of off-resonance coupling on the MS and Cardioid(2,3) gates. As can be seen, in the MS gate fast oscillations of SD and DS populations are caused by off-resonance carrier coupling. These are significantly suppressed in the Cardioid gate.
Intuitively this works since a smoother gate envelope has a smaller high frequency content and therefore less overlap with off-resonance transitions such as the carrier. Using smooth gate envelopes to reduce carrier coupling has already been suggested \cite{Roos2008} and implemented \cite{Kirchmair2009}. Here we have provided a general treatment that does not rely on spectral density arguments and does not require intermediate additional pulses as proposed in \cite{Roos2008}, which compensate for a spectral overlap with the red and blue side-band transitions. 

As a final demonstration we optimize the gate to mitigate normal mode frequency errors. Here, we set $\nu=\nu_0+\delta\nu$ and pursuit a path similar to that of the gate time error. This yields a set of constraints $V_{\boldsymbol{n^{-1}}}\boldsymbol{r}=\boldsymbol{0}$, where this time the Vandermonde matrix is $\left(V_{\boldsymbol{n}^{-1}}\right)_{i,j}=\left(n_{j}\right)^{-i}$. As opposed to timing-errors robustness, the quadratic term cannot be eliminated and the fidelity always scales quadratically in $\frac{\delta\nu}{\xi_0}$. However, the pre-factor of this quadratic term is minimized. The gate purity, which is a measure of disentanglement from the motion, given by $P_g\equiv\text{Tr}\left(\hat{\rho}^2\right)$, does become robust order-by-order. This is important since it reduces the number of error syndromes that need to be considered when implementing quantum error correction. This quadratic scaling of the fidelity encourages us to add only one degree of freedom as minimizing higher orders will have a marginal effect. However adding further harmonic tones may be used to combine this with eliminating timing errors and off-resonance carrier coupling. We denote this type of entangling combined scheme as a \textit{CarNu gate}. Fig. \ref{FigPhase}, \ref{FigTime}d and \ref{FigFidel}b show the phase space trajectory, populations evolution and fidelity, respectively, of the CarNu(2,3,7) gate which demonstrates robustness to gate timing-errors, off-resonance carrier coupling as well as normal mode frequency error.

In conclusion, we have analytically derived and experimentally demonstrated a scheme for robust entanglement gates for trapped-ion qubits. Our scheme increases the robustness to gate timing-errors and normal mode frequency errors as well as reduces off-resonance carrier coupling. This allows for the use of higher laser power and implementation of faster entangling gates while maintaining high gate fidelities. This optimization is particularly important when working with larger Coulomb crystals where the spectral distance between modes is small. From a broader point of view, we believe our methodology offers a simple and straight-forward prescription for increasing the efficiency of entangling operations, which are an essential tool in quantum information experiments as well as many other research directions.

\begin{acknowledgments}
This work was supported by the Crown Photonics Center, ICore-Israeli excellence center circle of light,
the Israeli Science Foundation, the Israeli Ministry of Science Technology and Space, the Minerva Stiftung
and the European Research Council (consolidator grant 616919-Ionology).
\end{acknowledgments}

\begin{center}
\textbf{\large Supplementary Materials}
\end{center}

\section{1. Energy levels, drive scheme and evolution of MS gate generalization}
\begin{figure}
\includegraphics[width=8.5cm]{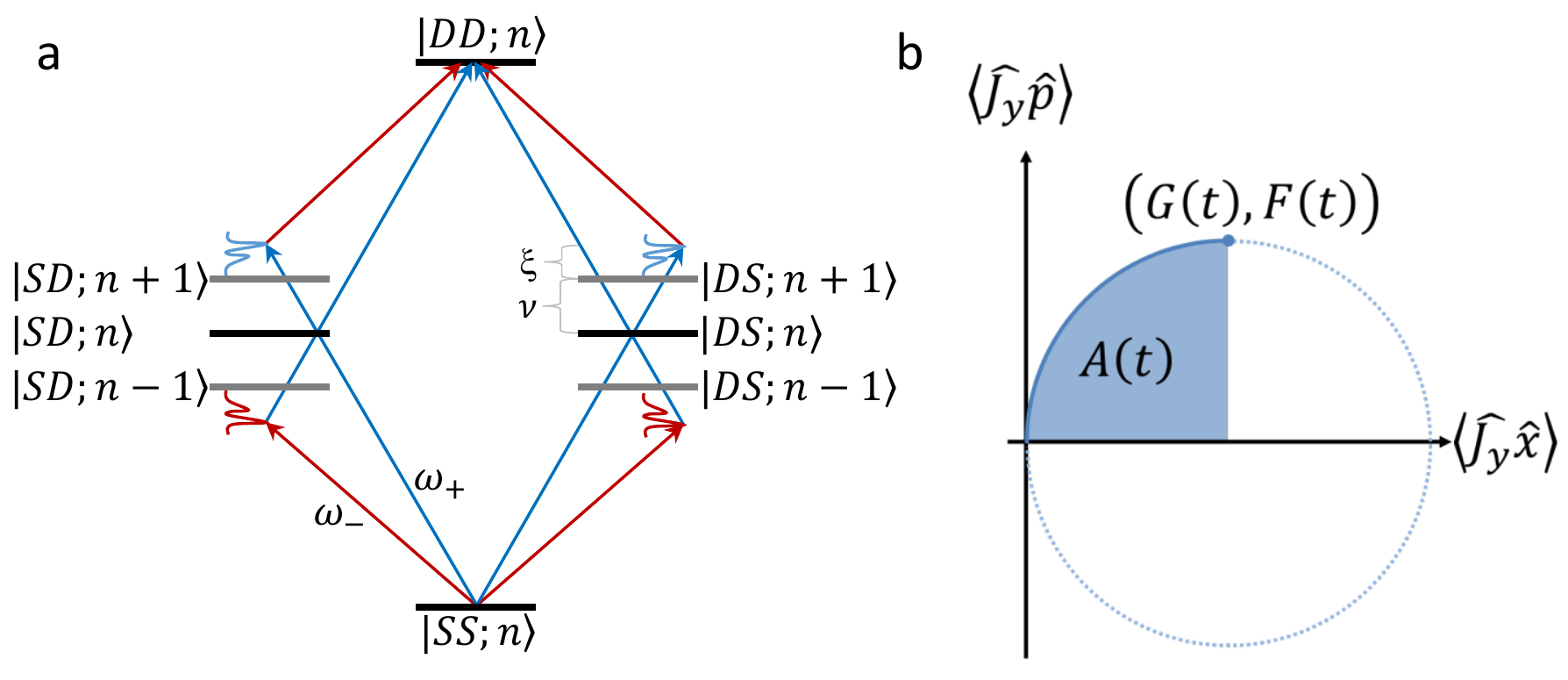}
\caption{\textbf{Energy levels, gate scheme and evolution.} \textbf{(a)} The system consists of two effective two level systems that constitute the four states SS,SD,DS and DD which are superimposed by a harmonic normal mode, characterized by the phonon number $n$ and frequency $\nu$. The MS gate uses a bi-chormatic laser field (red and blue arrows) which are detuned by $\xi_0$ symmetrically from the normal mode side-bands. For the generalized scheme we use a multi-chromatic laser field (represented by the red and blue two-peak spectra), which consists of harmonic tones of the frequency $\xi_0$ that are then superimposed onto the normal mode side-bands symmetrically. \textbf{(b)} The normal mode evolution forms a trajectory on the normal mode phase space, multiplied by the eigenvalue of $\hat{J}_{y}$, with coordinates $\left(G\left(t\right),F\left(t\right)\right)$. The area enclosed by this trajectory corresponds to a correlated rotation angle, $A\left(t\right)$, in the two-qubit sub-space, here we show the trajectory of the MS gate, which corresponds to $N=1$ and $n_1=1$ of the generalized gate.}\label{FigH}
\end{figure}

As described in the main text,  the MS gate is implemented by driving the qubit transition in a two-ion crystal with a bi-chromatic field containing the frequencies $\omega_{\pm}=\omega_{0}\pm\left(\nu+\xi_0\right)$. $\omega_{0}$ is the resonance frequency of the $\ensuremath{\left|S\right\rangle }\leftrightarrow\ensuremath{\left|D\right\rangle }$ transition, $\nu$ is the harmonic trap frequency of a selected normal mode of the crystal and $\xi_0$ is a frequency detuning from the side-band transition. Similarly, the generalized entangling gate is implemented by driving the qubit transition in a two-ion crystal with a multi-chromatic field containing frequency pairs $\omega_{\pm,i}=\omega_{0}\pm\left(\nu+n_i\xi_0\right)$. Fig. \ref{FigH}(a) shows schematically the energy levels of the two-qubit Hilbert space and the coupling fields for the MS gate (arrows) and generalized gate (spectrum-like lines). The corresponding unitary evolution is a trajectory $\left(G\left(t\right),F\left(t\right)\right)$ on the normal mode phase space, which is accompanied by a correlated rotation in the two-qubit subspace by an angle $A\left(t\right)$, which corresponds to the area enclosed by the trajectory. Fig. \ref{FigH}(b) shows the circular trajectory of the MS gate.

\section{2. Analytic expression for $\text{Cardioid}\left(1,2,...,N\right)$ gate}
\begin{figure}\includegraphics[width=8.5cm]{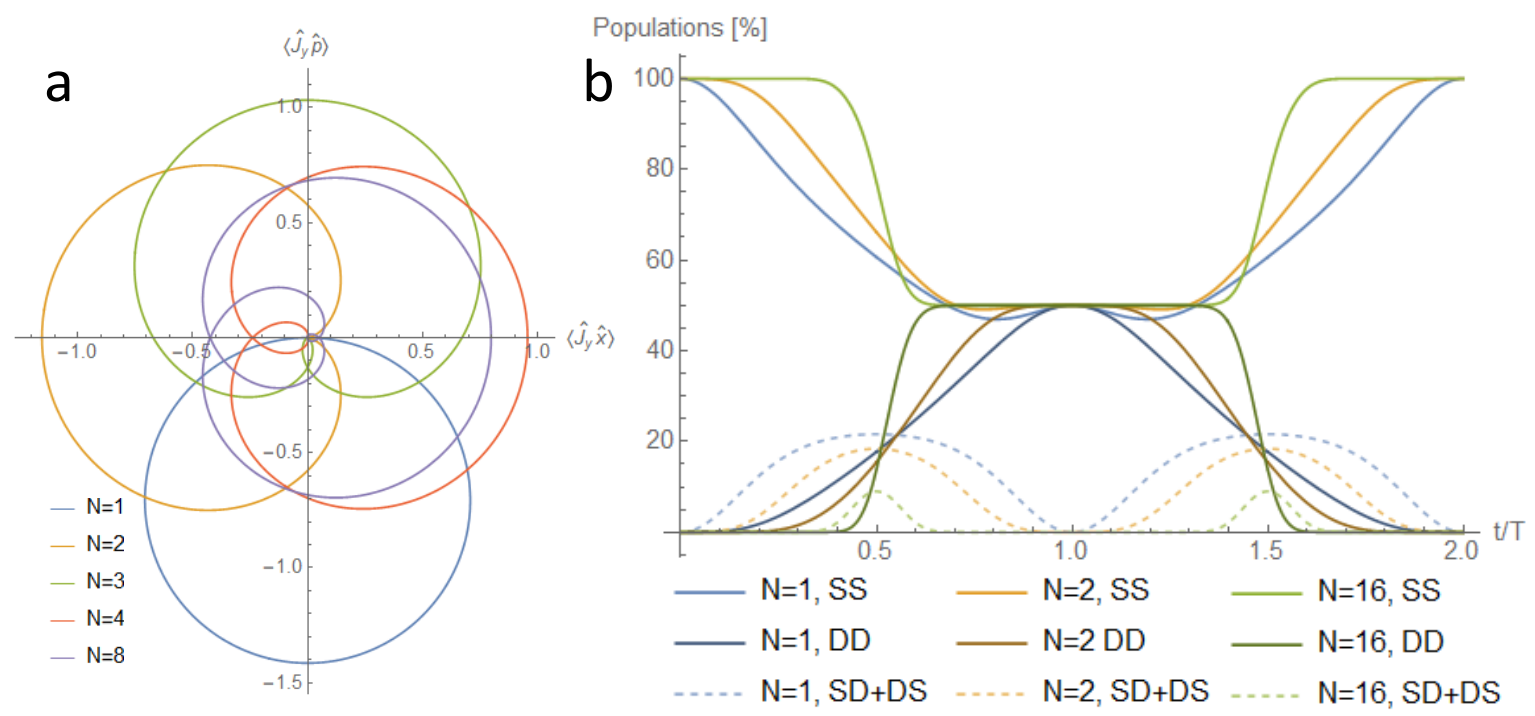}
\caption{\textbf{Phase space trajectory and gate population evolution of $\text{Cardiod}\left(1,2,...,N\right)$.} \textbf{(a)} Phase space trajectories. As the number of tones ($N$) increases, the trajectory becomes more centered around the origin, thus exciting less motion. \textbf{(b)} Gate evolution for $N=1,2,8$. As $N$ increases the maximum population of the unwanted SD and DS states decreases, and they decay fast to zero around the gate time. Accordingly the populations of SS and DD quickly approach 50\% and remain flat around the gate time.}\label{FigCarN}\end{figure}

The constraints required for a Cardioid(1,2,...,N) gate are $V_{\boldsymbol{n}}\boldsymbol{r}=\boldsymbol{0}$ where $\boldsymbol{r}$ is a vector of $N$ amplitudes and $V_{\boldsymbol{n}}$ is a $N\times N$ Vandermonde matrix defined by $\left(V_{\boldsymbol{n}}\right)_{i,j}=\left(j\right)^{i-1}$. The constraints are satisfied by the amplitudes:
\begin{equation}
r_{j}=\left(-1\right)^{N-j}\frac{N!}{2^{N}}\sqrt{\frac{2\sqrt{\pi}}{\left(N-1\right)!\Gamma\left(N+\frac{1}{2}\right)}}{N-1 \choose j-1}.
\end{equation}
Setting this in the expression for $F$ and $G$ we obtain a complicated expression. However it becomes more convenient in radial coordinates:
\begin{equation}
\begin{cases}
R\left(t\right)=\sqrt{G^{2}\left(t\right)+F^{2}\left(t\right)}=\sqrt{\frac{\sqrt{\pi}\left(N-1\right)!}{\Gamma\left(N+\frac{1}{2}\right)}}\left|\sin^{N}\left(\frac{\xi_{0}t}{2}\right)\right|\\
\phi\left(t\right)=\arctan\left(\frac{F\left(t\right)}{G\left(t\right)}\right)=\left(-1\right)^{N}\frac{N\xi_{0}t}{2}
\end{cases}.
\end{equation}
Where the pre-factor for $R$ scales as $\left(\frac{\pi}{N}\right)^{\frac{1}{4}}$ for $N\gg1$. The correlated rotation angle is given by:
\begin{equation}
A\left(t\right)=a\left(t\right)-\frac{1}{4}R^{2}\left(t\right)\sin\left(2\phi\left(t\right)\right).
\end{equation}
With:
\begin{equation}
a\left(t\right)=\frac{\pi}{4}-\frac{\sqrt{\pi}N!\cos\left(\frac{\xi_{0}t}{2}\right)}{2\Gamma\left(N+\frac{1}{2}\right)}\,_{2}F_{1}\left(\frac{1}{2},\frac{1}{2}-N;\frac{3}{2};\cos^{2}\left(\frac{\xi_{0}t}{2}\right)\right).
\end{equation}
Where $\,_{2}F_{1}\left(a,b;c;z\right)$ is the hyper-geometric function. This allows for a closed form of the Cardioid(1,2,...,N) evolution and fidelity:
\begin{equation}
\begin{cases}
Pr\left(SS\right)=\frac{3+e^{-4\left(\bar{n}+\frac{1}{2}\right)\frac{R^{2}\left(t\right)}{2}}+4e^{-\left(\bar{n}+\frac{1}{2}\right)\frac{R^{2}\left(t\right)}{2}}\cos\left(a\left(t\right)\right)}{8}\\
Pr\left(DD\right)=\frac{3+e^{-4\left(\bar{n}+\frac{1}{2}\right)\frac{R^{2}\left(t\right)}{2}}-4e^{-\left(\bar{n}+\frac{1}{2}\right)\frac{R^{2}\left(t\right)}{2}}\cos\left(a\left(t\right)\right)}{8}\\
Pr\left(SD\right)=\frac{1-e^{-4\left(\bar{n}+\frac{1}{2}\right)\frac{R^{2}\left(t\right)}{2}}}{8}\\
Pr\left(DS\right)=\frac{1-e^{-4\left(\bar{n}+\frac{1}{2}\right)\frac{R^{2}\left(t\right)}{2}}}{8}\\
F_{g}^{2}=\frac{3+e^{-4\left(\bar{n}+\frac{1}{2}\right)\frac{R^{2}\left(t\right)}{2}}}{8}+\frac{\sin\left(a\left(t\right)\right)e^{-\frac{R^{2}\left(t\right)}{2}\left(\bar{n}+\frac{1}{2}\right)}}{2}
\end{cases}.
\end{equation}
Fig. \ref{FigCarN} shows the phase space trajectories and evolution of different Cardioid gates of this form, clearly using more tones leads to a flatter response around the gate time $T$. We also note that assuming we are constrained by the total Rabi frequency, which corresponds to a MS gate time $T_{\text{MS}}$, then the gate time scales linearly with the number of tones: $T_{\Omega}=T_{\text{MS}}\frac{N^{2}}{2N-1}\stackrel{N\gg1}{\longrightarrow}T_{\text{MS}}\frac{N}{2}$.

\section{3. Active non-linear compensation and $\text{Cardioid}\left(1,2\right)$}
\begin{figure}\includegraphics[width=8.5cm]{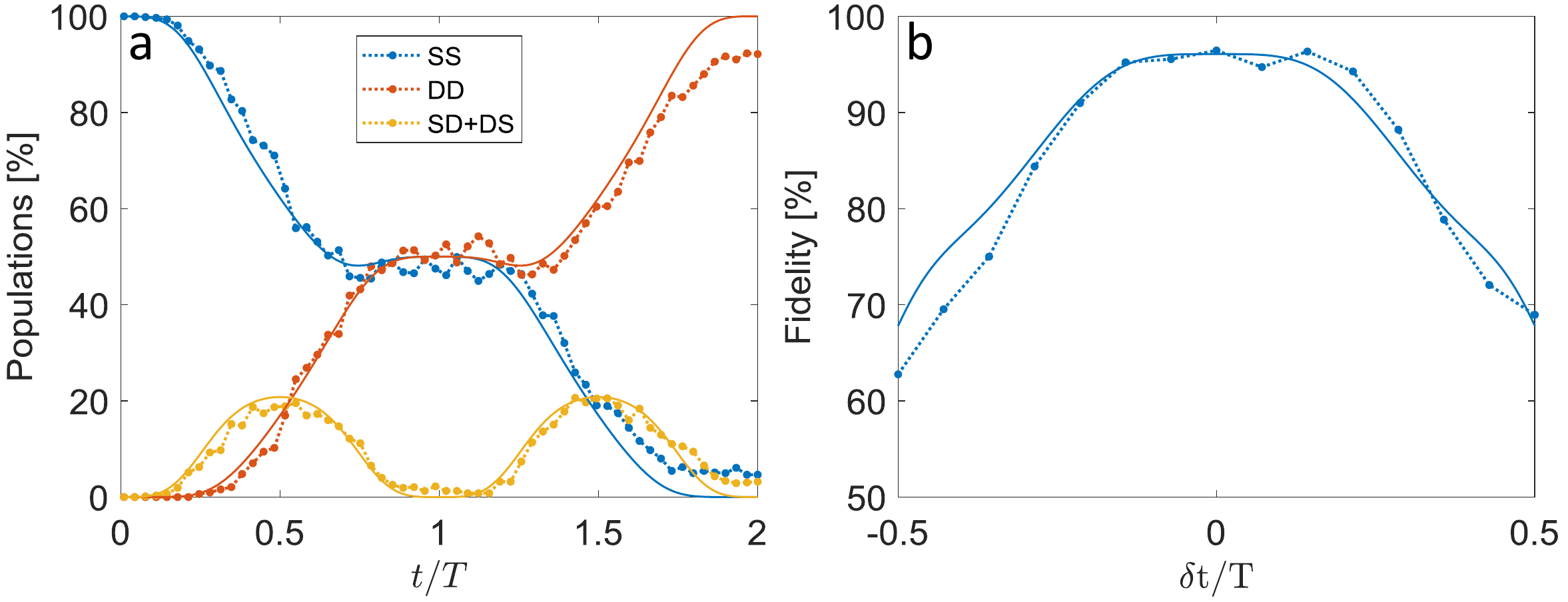}
\caption{\textbf{Cardioid(1,2) using active non-linear compensation method.} All data (points connected by dashed line) are obtained by averaging 625 realizations yielding $2\%$ projection error at $50\%$ probability. \textbf{(a)} Population evolution of a Cardioid(1,2) with active compensation of non-linear response. Analytic solution (solid lines) are without fitting parameters. The gate evolution is well behaved and fits the theory, nevertheless some decoherence-like effect are seen around $t=2T$ as the SD and DS populations do not return to 0. \textbf{(b)} Cardioid(1,2) robustness to gate timing-errors. Fit (solid line) is obtained by rescaling and shifting the analytic solution. The fidelity flattens around the gate time.}\label{FigCar12}\end{figure}

As described in the main text, when implementing schemes as Cardioid(1,2) or CarNu(1,2,3) etc. a third order non-linear response of different components result in unwanted on-resonance side-band excitations. To overcome this we used sequences which do not create such a response, these are sequences where a sum or difference of three tones cannot be $0$. However, implementing a $n_i=\left\{1,2,3,...\right\}$ sequence is still possible by using active compensation to negate the unwanted non-linear response - that is, in addition to the signals required by the entangling scheme we superimpose tones which destructively interfere with the unwanted on-resonance components. Fig. \ref{FigCar12} shows the evolution and gate time error fidelity of the Cardioid(1,2) scheme. We note that this active compensation approach requires a feedback loop due to the instability of the non-linear responses and is thus not preferable.

\end{document}